\documentclass[%
 aip,
rsi,%
 amsmath,amssymb,
 reprint,%
]{revtex4-1}

\usepackage{xcolor}
\usepackage{graphicx}% Include figure files
\usepackage{dcolumn}% Align table columns on decimal point
\usepackage{bm}% bold math
\usepackage{nicefrac}
\newcommand{\n}[1]{\mathbf{#1}}

\begin{document}

\preprint{AIP/123-QED}

\title[Rodr\'{i}guez et al.]{Necessary and Sufficient Conditions for Quasisymmetry}% Force line breaks with \\

\author{E. Rodr\'{i}guez}
 \altaffiliation[Email: ]{eduardor@princeton.edu}%Lines break automatically or can be forced with \\
 \affiliation{ 
Department of Astrophysical Sciences, Princeton University, Princeton, NJ, 08543%\\This line break forced with \textbackslash\textbackslash
}
\affiliation{%
Princeton Plasma Physics Laboratory, Princeton, NJ, 08540%\\This line break forced% with \\
}%

\author{P. Helander}%
\altaffiliation[Email: ]{per.helander@ipp.mpg.de}%
\affiliation{%
Max Planck Institute for Plasma Physics, Wendelsteinstrasse 1, 17491 Greifswald, Germany%\\This line break forced% with \\
} 

\author{A. Bhattacharjee}
 \altaffiliation[Email: ]{amitava@princeton.edu}
 \affiliation{ 
Department of Astrophysical Sciences, Princeton University, Princeton, NJ, 08543%\\This line break forced with \textbackslash\textbackslash
}
\affiliation{%
Princeton Plasma Physics Laboratory, Princeton, NJ, 08540%\\This line break forced% with \\
}%

\date{\today}% It is always \today, today,

\begin{abstract}
A necessary and sufficient set of conditions for a quasisymmetric magnetic field in the form of constraint equations is derived from first principles. Without any assumption regarding the magnetohydrodynamic (MHD) equilibrium of the plasma, conditions for quasisymmetry are constructed starting from the single-particle Lagrangian to leading order. The conditions presented in the paper are less restrictive than the set recently obtained by Burby et al. [J.  W.  Burby,  N.  Kallinikos,  R.  S.  MacKay, arXiv:1912.06468 (2019)], and could facilitate ongoing efforts towards investigating the existence of global quasisymmetric MHD equilibria. It is also shown that quasisymmetry implies the existence of flux surfaces, regardless of whether the field corresponds to an MHD equilibrium.
\end{abstract}

\maketitle

\section{\label{sec:intro} Introduction:}

Stellarators, as well as other magnetic confinement devices such as tokamaks, serve the main purpose of containing a hot plasma and thermally insulating it from the surroundings. To do so, it is necessary that the great majority of all collisionless orbits are confined within the plasma volume. Of course, charged particles tend to move parallel to magnetic field lines while gyrating in the perpendicular direction, but they also drift perpendicularly to the field lines. 
 \par
 To prevent this drift from carrying particles out of the plasma, it is helpful to endow the magnetic geometry with a continuous symmetry. By Noether's theorem, this ensures the existence of a conserved quantity and, frequently, good orbit confinement. In the case of an axisymmetric tokamak, particles must conserve their \textit{canonical momentum}, which forces them to stay close to magnetic flux surfaces tangential to the $\n{B}$ field (Tamm's theorem).  \par
Stellarators, however, generally lack the spatial symmetry that characterises tokamaks, and therefore also this conserved momentum. Instead, they may benefit from so-called quasisymmetry (QS), which is a non-trival extension of the concepts of axisymmetry and helical symmetry. Due to the three-dimensional (3D) nature of stellarators, QS can be achieved \textit{without} imposing any continuous spatial symmetry on $\n{B}$. As shown in many works\cite{nuhren1988,boozer1995,Helander2014}, a conserved quantity arises in the lowest-order equations of guiding-center motion if the magnitude of the magnetic field $\n{B}$ in magnetohydrodynamic (MHD) equlibrium is independent of a linear combination of Boozer angular coordinates\cite{boozer1981,Helander2014}. It seems unlikely that this condition can be achieved exactly everywhere within the plasma (this remains to be proved rigorously), but in practice it is sufficient that it is attained to a good approximation.\cite{garbooz1991ne,plunk2018} 
\par
Recently, Burby et al.\cite{burby2019} used differential forms to derive conditions for the existence of QS, detached from any assumption on whether the field satisfies MHD force balance or not. In the present paper we consider the problem of deriving conditions for QS in this context \textit{without} the use of differential forms, but rather using standard vector analysis. This proves to be both simple and instructive, and also leads to a simplification of the conditions in [\onlinecite{burby2019}]. \par
The article is structured as follows. In Sec. II we first briefly introduce the leading-order single-particle Lagrangian describing the motion of charged particles. In Sec. III Noether's theorem is applied both to obtain the conditions for quasisymmetry as well as the associated conserved momentum. Sec. IV compares our results to other accounts of QS, in particular to [\onlinecite{burby2019}], and discusses them from a physical point of view.

\section{Single-particle Lagrangian}
Before dealing with QS, we introduce the Lagrangian for charged-particle motion in a magnetic field. This will be the basis on which the derivation of the QS conditions will be built, and therefore special attention should be paid to the assumptions underlying it. We construct an approximate form of the single-particle Lagrangian following Littlejohn\cite{littlejohn1983}, omitting some of the algebraic details. \par
The motion of a charged particle in a magnetic field is described by the Hamiltonian
\begin{equation}
    \mathcal{H}(\n{q},\n{p},t)=\frac{1}{2}\left[\mathbf{p}-\frac{1}{\epsilon}\mathbf{A}(\mathbf{q},\epsilon t)\right]^2, \label{eq:hamiltonian}
\end{equation}
where $\n{A}$ is the magnetic vector potential $\n{B}=\nabla\times\n{A}$, we have ignored the electric field. and natural units ($e=m=c=1$) have been used. The parameter $\epsilon$ serves as an ordering parameter, quantifying the strength of the magnetic field and thus the validity of the guiding-center approximation of particle motion. An order-of-magnitude estimate shows that $A\sim LB$, where $L$ is some characteristic field length scale, and $p\sim \rho B$ where $p$ is the magnitude of the particle momentum and $\rho$ is the gyroradius of the particle. Thus, one may understand the parameter $\epsilon\sim\rho/L$, which we will assume to be small. Note that the time variation of the vector potential $\n{A}$ is assumed to be slow, and accordingly taken to be $O(\epsilon$). \par
The Lagrangian is constructed from the Hamiltonian (\ref{eq:hamiltonian}) by applying an appropriate Lagrange transformation. First, we need to define the Lagrangian variables,
\begin{align}
    \mathbf{x}=\mathbf{q} \\
    \mathbf{v}=\mathbf{p}-\frac{1}{\epsilon}\mathbf{A}(\mathbf{q},\epsilon t).
\end{align}
Then,
\begin{align}
    \mathcal{L}(\mathbf{x},\n{v},\mathbf{\Dot{x}},t)=\left[\frac{1}{\epsilon}\mathbf{A}(\mathbf{x},\epsilon t)+\mathbf{v}\right]\cdot\mathbf{\Dot{x}}-\frac{v^2}{2}. \label{eqn:lagrangian}
\end{align}
To evaluate the Lagrangian $\mathcal{L}$ order by order, we separate the velocity $\mathbf{v}=v_\parallel\mathbf{b}+v_\perp\mathbf{c}$, where $\mathbf{b}$ is the unit vector along the magnetic field and $\mathbf{c}\cdot\mathbf{b}=0$. Let us further introduce the gyrocentre position $\mathbf{x}=\mathbf{X}+\epsilon v_\perp\mathbf{a}/B$, where $\mathbf{a}=\cos\theta\mathbf{e_1}-\sin\theta\mathbf{e}_2$ represents the rotating gyromotion of the charged particles (perpendicular to $\n{c}$), $\theta$ is the gyrophase, and $\{\mathbf{e}_i\}$ denotes an orthonormal basis for the plane perpendicular to $\mathbf{b}$. \par
With this new set of variables we expand all terms in (\ref{eqn:lagrangian}) to $O(\epsilon)$. Dropping the explicit time dependence for simplicity, and simplifying the resulting expression by adding an appropriate total time derivative, we obtain\cite{littlejohn1983} 
\begin{equation*}
    \mathcal{L}^{(1)}(\mathbf{X},v_\parallel,\theta,v_\perp)=\left[\frac{1}{\epsilon}\mathbf{A}(\mathbf{X})+v_\parallel\mathbf{b}\right]\cdot\mathbf{\Dot{X}}-\frac{v_\parallel^2+v_\perp^2}{2}+O(\epsilon).
\end{equation*}
The term involving $v_\perp^2$ can be written in a more convenient form by invoking the conservation of the magnetic moment $\mu=v_\perp^2/2B$ (which one could obtain as an Euler-Lagrange equation for the ignorable coordinate $\theta$ from $\mathcal{L}^{(2)}$, see [\onlinecite{littlejohn1983}] for further details).
Taking $\mu$ to be an external variable, the final form of the single particle Lagrangian to $O(\epsilon)$ becomes
\begin{equation}
    \mathcal{L}^{(1)}(\mathbf{X},v_\parallel,\theta)=\left[\frac{1}{\epsilon}\mathbf{A}(\mathbf{X})+v_\parallel\mathbf{b}\right]\cdot\mathbf{\Dot{X}}-\frac{v_\parallel^2}{2}-\mu B+O(\epsilon). \label{eqn:lLag}
\end{equation}
The dynamics of $\mathbf{X}$ and $v_\parallel$ are then described by the Euler-Lagrange equations associated to $\mathcal{L}^{(1)}$. Proceeding perturbatively order by order, we find
\begin{align}
    \mathbf{\Dot{X}}=&[ v_\parallel+O(\epsilon)]\mathbf{b}+\nonumber \\ &+\epsilon\left[\frac{v_\parallel^2}{B}\n{b}\times(\n{b}\cdot\nabla\n{b})+\frac{\mu}{B}\n{b}\times\nabla B\right]+O(\epsilon^2) \label{eqn:dX}\\
    \dot{v}_\parallel=&-\mu\mathbf{b}\cdot\nabla B+O(\epsilon) .
\end{align}
As expected, Eq. (\ref{eqn:dX}) includes the usual gradient and curvature drifts at order $\epsilon$.\cite{Alfven1940} The dynamics is largely determined by the scalars $B$ (through the magnetic moment $\mu$ conservation) and $\n{b}\cdot\nabla B$, which will later prove to be important.
\par
Of course, none of the material in this section is new. It is however included as an introduction of our notation and, more importantly, as a reminder of the level of accuracy of the construction.
%%%%%%%%%%%%%%%%%%%%%%%%%%%%%%%%%%%%%%%%%%%%%%%%%
\section{Quasisymmetry constraint equations}
Given the leading-order Lagrangian in its form (\ref{eqn:lLag}), we are now in a position to look for symmetries that will lead to conserved momenta. \par
In the language of Noether's theorem, the search for the existence of a spatial symmetry is equivalent to the search for some coordinate $\Bar{q}=\Bar{q}(\n{X})$ such that
$$ \partial_{\Bar{q}}\mathcal{L}^{(1)}|_{\n{z}\neq \Bar{q},\Dot{\n{z}}}=0, $$
where $\n{z}$ refers to the whole set of Lagrangian coordinates. Provided this holds true, there will be an associated conserved conjugate momentum $\Bar{p}=\partial_{\Dot{\Bar{q}}}\mathcal{L}^{(1)}$, which would restrict the motion of particles in real space. \par
We thus consider the derivative of the Lagrangian, and apply the chain rule, dropping any explicit time dependence,
\begin{align*}
    \frac{\partial\mathcal{L}^{(1)}}{\partial \Bar{q}}=&\frac{\partial\mathcal{L}^{(1)}}{\partial \n{X}}\cdot\frac{\partial\n{X}}{\partial \Bar{q}}+\frac{\partial\mathcal{L}^{(1)}}{\partial \n{\Dot{X}}}\cdot\frac{\partial\n{\Dot{X}}}{\partial \Bar{q}}=\\
    =& \frac{\partial\mathcal{L}^{(1)}}{\partial \n{X}}\cdot\frac{\partial\n{X}}{\partial \Bar{q}}+\frac{\partial\mathcal{L}^{(1)}}{\partial \n{\Dot{X}}}\cdot\frac{\mathrm{d}}{\mathrm{d}t}\left(\frac{\partial\n{X}}{\partial \Bar{q}}\right) \\
    =& \frac{\partial\mathcal{L}^{(1)}}{\partial \n{X}}\cdot\frac{\partial\n{X}}{\partial \Bar{q}}+\frac{\partial\mathcal{L}^{(1)}}{\partial \n{\Dot{X}}}\cdot(\n{\Dot{X}}\cdot\nabla)\frac{\partial\n{X}}{\partial \Bar{q}}.
\end{align*}
In terms of the symmetry vector field $\n{u}\stackrel{.}{=}\partial\n{X}/\partial \Bar{q}$, the symmetry condition may now be written in the following form, 
\begin{equation}
    0=\n{u}\cdot\partial_{\n{X}}\mathcal{L}^{(1)}+\partial_{\n{\Dot{X}}}\mathcal{L}^{(1)}\cdot(\n{\Dot{X}}\cdot\nabla)\n{u},
\end{equation}
where, according to (\ref{eqn:lLag}),
\begin{align*}
    \partial_{\n{X}}\mathcal{L}^{(1)}=&\frac{1}{\epsilon}(\nabla\n{A})\cdot\n{\Dot{X}}+v_\parallel(\nabla\n{b})\cdot\n{\Dot{X}}-\mu\nabla B \\
    \partial_\n{\Dot{X}}\mathcal{L}^{(1)}=&\frac{1}{\epsilon}\n{A}+v_\parallel\n{b}.
\end{align*}
Collecting terms, we obtain
\begin{align}
    0=&\frac{1}{\epsilon}\left(\n{u}\cdot\nabla\n{A}\cdot\n{\Dot{X}}+\n{\Dot{X}}\cdot\nabla\n{u}\cdot\n{A}\right)+\nonumber\\
    &+v_\parallel\left(\n{u}\cdot\nabla\n{b}\cdot\n{\Dot{X}}+\n{\Dot{X}}\cdot\nabla\n{u}\cdot\n{b}\right)-\mu\n{u}\cdot\nabla B. \label{lagureq}
\end{align}
In order for QS to be a symmetry that holds for any charged particle in the system, the equality in (\ref{lagureq}) should hold for all $\mu$ and $v_\parallel$. Equally importantly, we remark that the Lagrangian $\mathcal{L}^{(1)}$ only holds approximately up to $O(\epsilon)$, and thus so should (\ref{lagureq}). This implies that each of the three clusters in (\ref{lagureq}) must vanish separately to $O(\epsilon)$. \par
Let us start with the simplest, final term in (\ref{lagureq}), which gives the first of three QS conditions,
\begin{equation}
    \n{u}\cdot\nabla B=0,
\end{equation}
and thus implies that the magnetic field strength must be uniform in the symmetry direction $\n{u}$. 
\par
Let us now focus on the first term in (\ref{lagureq}), which is larger than the others by a factor of $\epsilon^{-1}$. This term means that particles must follow field lines to lowest order in $\epsilon$, and  gives the condition
\begin{align}
    \frac{1}{\epsilon}&\left(\n{u}\cdot\nabla\n{A}\cdot\n{\Dot{X}}+\n{\Dot{X}}\cdot\nabla\n{u}\cdot\n{A}\right)=\nonumber\\
    &=\frac{1}{\epsilon}\n{\Dot{X}}\cdot\left(\n{u}\cdot\nabla\n{A}+\n{A}\cdot\nabla\n{u}+\n{A}\times\nabla\times\n{u}\right)= \nonumber \\
    &=\frac{1}{\epsilon}\n{\Dot{X}}\cdot\left[-\n{u}\times\n{B}+\nabla(\n{A}\cdot\n{u})\right]=0. \label{eqn:symAterm}
\end{align}
To order $1/\epsilon$, and according to the equation of motion (\ref{eqn:dX}), the $\n{b}$ projection of the expression in square brackets in (\ref{eqn:symAterm}) must vanish. As we expect the expressions to hold up to, but not including, order $\epsilon$, the $\n{\Dot{X}}^{(1)}$ component of the square bracket must also vanish. The perpendicular drift components of (\ref{eqn:dX}) were shown to be along $\n{b}\times(\n{b}\cdot\nabla\n{b})$ and $\n{b}\times\nabla B$, which generally are not collinear. (They are parallel in a force-free field, which we treat later in this section.)\footnote{Note that more generally any field with $\n{j}=\lambda\n{B}+\alpha\n{b}\times\nabla B$ (where $\lambda$ and $\alpha$ represent arbitrary functions) will have colinear magnetic gradient and curvature drifts. This class includes the force-free case, which we made explicit reference to in the text due to its particular relevance. However, Eq.~(\ref{forceFree}) will hold not just for the force-free, but for the whole class.} Because the different terms in (\ref{eqn:symAterm}) depending on $v_\parallel$ and $\mu$ must vanish separately, three independent projections of the whole expression inside the square bracket vanish, and thus so must the bracket itself. \par
Let us define $\n{A}\cdot\n{u}\stackrel{.}{=}-\psi$, and adopt, for instance, the Coulomb gauge $$ \n{A}(\n{r})=\int\frac{\n{j}(\n{r}')}{|\n{r}-\n{r}'|}\mathrm{d}^3\n{r}'.$$ This integral form is single-valued and converges provided the current density is well-behaved and has a finite support. With this single-valued definition of $\psi$, we obtain the second QS condition,
% If we define $\n{A}\cdot\n{u}\stackrel{.}{=}-\psi$, and adopt, for example, the Coulomb gauge for the vector potential, so that $\psi$ is a single-valued function}, we obtain the second QS condition
\begin{equation}
    \n{B}\times\n{u}=\nabla\psi ~~\implies~~ \nabla\times(\n{B}\times\n{u})=0. \label{C2or}
\end{equation}
Finally, we consider the term in (\ref{lagureq}) that contains $v_\parallel$. One may straightforwardly write
\begin{equation}
    v_\parallel\n{\Dot{X}}\cdot\left(\n{u}\cdot\nabla\n{b}+\nabla\n{u}\cdot\n{b}\right)=0. \label{eqn:symVterm}
\end{equation}
This condition must hold up to order $\epsilon$, which implies that only the $\n{b}$ component of the expression in the brackets must vanish, i.e.
\begin{align*}
    \n{b}\cdot\left(\n{u}\cdot\nabla\n{b}+\nabla\n{u}\cdot\n{b}\right)=\n{b}\cdot\nabla\n{u} \cdot \n{b}=0,
\end{align*}
where the first term vanishes because $\n{b}$ is a unit vector. 
This condition could be recast into a simpler form using (\ref{C2or}),
\begin{align*}
    \nabla&\times(\n{u}\times\n{B})=\\
    &=\n{u}(\nabla\cdot\n{B})-\n{B}(\nabla\cdot\n{u})+(\n{B}\cdot\nabla)\n{u}-(\n{u}\cdot\nabla)\n{B}=0 \\
    &\implies\n{b}(\nabla\cdot\n{u})=(\n{b}\cdot\nabla)\n{u}-(\n{u}\cdot\nabla)\n{b}.
\end{align*}
Thus, the third QS condition is
\begin{equation}
    \n{b}\cdot\nabla\n{u} \cdot \n{b}=\nabla\cdot\n{u}+\n{b}\cdot(\n{u}\cdot\nabla)\n{b}=\nabla\cdot\n{u}=0
\end{equation}
and we conclude that the vector field $\n{u}$ must be divergence-free. \par
Collecting the three conditions above, for a magnetic field configuration to be quasisymmetric, it is necessary that there exists a vector field $\n{u}$ such that the following set of equations is satisfied,
\begin{subequations}
\begin{align}
    \n{u}\cdot\nabla B=0 \label{C1}\\
    \n{B}\times\n{u}=\nabla\psi \label{C2} \\
    \nabla\cdot\n{u}=0, \label{C3}
\end{align}
\end{subequations}
where $\psi=-\n{A}\cdot\n{u}$. A more rigorous procedure within this formalism shows that $\psi$ may in fact be taken to be any single-valued function. The proof is omitted here. Interestingly, requiring QS implies the existence of magnetic flux surfaces that must be toroidal by the "hairy-ball" theorem. This is a direct consequence of $\mathbf{B}\cdot\nabla\psi=0$, with $\psi$ a single-valued function (with $\nabla\psi\neq0$ generally) and $\mathbf{B}$ non-vanishing. The symmetry alone, without any additional requirement on plasma equilibrium, leads to this conclusion. In addition, and because generally $B\neq B(\psi)$,\cite{bernardin1986} Eqn. (\ref{C1}) implies that $\n{u}$ is integrable.  In other words, since it is generally impossible that the field strength is constant on a magnetic surface, the streamlines of the vector field $\n{u}$ cannot trace out such surfaces but must close on themselves. As is well known, there are thus three topological possiblities of QS: poloidal, toroidal and helical.
\par
Let us now briefly consider the case in which the field is force-free, $(\nabla \times {\bf B}) \times {\bf B} = 0$, so that the vector $\n{\Dot{X}}$ lies in the plane spanned by the vectors $\mathbf{B}$ and $\mathbf{B} \times \nabla B$. Then one cannot conclude that the final bracket in Eq.~(\ref{eqn:symAterm}) must vanish, only that it must be parallel to $\nabla_\perp B = \nabla B - \mathbf{bb} \cdot \nabla B$. We then have
\begin{equation} 
\n{B}\times\n{u} = \nabla \psi + \lambda \nabla_\perp B, \label{forceFree}
\end{equation}
where $\lambda$ is some scalar function of position. Taking the scalar product with $\bf B$, we again conclude that ${\bf B} \cdot \nabla \psi = 0$ and that flux surfaces exist.
\par
Finally, let us find an expression for the conserved momentum of the charged particles $\Bar{p}$ associated to $\bar{q}$. To do so, we need to express the Lagrangian $\mathcal{L}^{(1)}$ in terms of a complete set of real-space coordinates $\{\Bar{q},q_2,q_3\}$ and introduce an orthogonal basis triad $\{\n{u},\Bar{\n{e}}_2,\Bar{\n{e}}_3\}$ related to these coordinates.
We define $q_1=\Bar{q}$, and $\Bar{\n{e}}_2$ and $\Bar{\n{e}}_3$ to be the corresponding unit vectors associated to two coordinates orthogonal to $\n{u}$. Finally we define the set so that $\Bar{\n{e}}_2\times\Bar{\n{e}}_3=\n{u}/|\n{u}|$. This definition of local coordinates is always possible provided $|\n{u}|\neq0$.\par
In this basis, one may express $\n{\Dot{X}}=\Dot{\Bar{q}}\n{u}+\Dot{q}_2\Bar{\n{e}}_2+\Dot{q}_3\Bar{\n{e}}_3$ and thus
\begin{align*}
    \mathcal{L}^{(1)}=&\frac{1}{\epsilon}\left[-\psi\Dot{\Bar{q}}+\Dot{q}_2\mathbf{A}\cdot\Bar{\n{e}}_2+\Dot{q}_3\mathbf{A}\cdot\Bar{\n{e}}_3\right]+\\
    &+v_\parallel\left(\Dot{\Bar{q}}\n{u}\cdot\n{b}+\Dot{q}_2\mathbf{b}\cdot\Bar{\n{e}}_2+\Dot{q}_3\mathbf{b}\cdot\Bar{\n{e}}_3\right)-\frac{v_\parallel^2}{2}-\mu B.
\end{align*}
Therefore, the conserved momenta of charged particles to $O(\epsilon)$ is
\begin{equation}
    \Bar{p}=\partial_{\Dot{\Bar{q}}}\mathcal{L}^{(1)}=-\frac{1}{\epsilon}\psi+v_\parallel\n{u}\cdot\n{b}. \label{momCons}
\end{equation}
This form of the conserved momentum is a natural generalisation of the conserved canonical momentum in axisymmetric systems. \par 
As a result of the conservation, the particle motion is restricted to $\psi$ surfaces more or less effectively depending on the particle parallel velocity. In fact, one could estimate the size of departure $\Lambda\sim\epsilon\rho_\parallel(B_\parallel/B_\perp)_\n{u}$ where $\rho_\parallel=v_\parallel/B$ is the `effective gyroradius' due to the parallel velocity, and $B_\parallel$ and $B_\perp$ correspond to projections of the $\n{B}$ field along and perpendicularly to $\n{u}$. One then expects enhanced neoclassical transport to occur when the symmetry is aligned with the $\n{B}$ field (as it happens to a large extent in a tokamak). Neoclassical transport would be suppressed to the level of classical transport if $\n{B}$ and $\n{u}$ were orthogonal. In quasi-helically and quasi-poloidally symmetric fields, the angle between $\n{u}$ and $\n{B}$ tends to be larger than in quasi-axisymmetric fields. This makes the ``banana'' orbit width smaller, reduces the bootstrap current, and improves fast-ion confinement.

\section{Alternative formulations}
So far, we have reduced the problem of quasisymmetry to the existence of a symmetry vector field $\n{u}$ satisfying the constraint equations (\ref{C1}) through (\ref{C3}). We now compare this formulation to other formulations of QS conditions in the literature. 

\subsection{Axisymmetric case}
Let us start by making contact with the simplest case of  QS, namely that of continuous axisymmetry. From the constraint equation (\ref{C2}), it follows that generally a QS magnetic field may be written in terms of the vector field $\n{u}$ as
\begin{equation}
    \n{B}=\frac{1}{|\n{u}|^2}(C\n{u}+\n{u}\times\nabla\psi), \label{genAS}
\end{equation}
taking $\n{u}\cdot\nabla\psi=0$ and $C=\n{B}\cdot\n{u}$. \par
If this form is equated to the axisymmetric (AS) one $\n{B}=F(R,z)\nabla\phi+\nabla\phi\times\nabla\psi(R,z)$ in standard cylindrical coordinates, one finds $\n{u}=R^2\nabla\phi$, with $C=F$. One may easily check that conditions (\ref{C1}) and (\ref{C3}) follow. \par 
The AS field is thus clearly a special case of the more general class of quasisymmetry. In fact, (\ref{genAS}) might be then interpreted as a generalised form of the axisymmetric field. For completeness, the canonical momentum might be written applying (\ref{momCons}) as $\Bar{p}=R(A_\phi/\epsilon+v_\phi)$, as expected.

\subsection{Triple product definition of quasisymmetry}
A particularly economical way of formulating QS is to use the following triple product as in (66) of [\onlinecite{Helander2014}],
\begin{equation}
    (\nabla\psi\times\nabla B)\cdot\nabla(\n{B}\cdot\nabla B)=0, \label{QSHel}
\end{equation}
where $\psi$ defines flux surfaces $\n{B}\cdot\nabla\psi=0$ (which are assumed to exist). We shall now prove that this expression is equivalent to the set of constraint equations (\ref{C1})-(\ref{C3}). \par
First, we construct $\n{u}$. From (\ref{C2}), and defining $u_\parallel\stackrel{.}{=}\n{b}\cdot\n{u}$ and $\n{u}_\perp\stackrel{.}{=}\n{u}-u_\parallel\n{b}$, we have
\begin{gather*}
    \n{u}_\perp=\frac{\nabla\psi\times \n{B}}{B^2}.
\end{gather*}
From (\ref{C1}), and assuming that $\n{B}\cdot\nabla B\neq0$,
\begin{gather*}
    u_\parallel= \frac{\n{b}\times\nabla\psi\cdot\nabla B}{\n{B}\cdot\nabla B},
\end{gather*}
which combines with $\n{u}_\perp$ to give
\begin{align}
    \n{u}=\frac{\nabla\psi\times\nabla B}{\n{B}\cdot\nabla B}. \label{eqn:uField}
\end{align}
Having constructed the symmetry vector field $\n{u}$, enforcing the third of the constraints (\ref{C3}) yields
precisely the triple product form (\ref{QSHel}) of QS. \par
It remains to be shown that one may enact the reverse, i.e., start from the triple product and the existence of flux surfaces to deduce the constraint equations.
To this end, define a vector field $\n{u}$ of the form (\ref{eqn:uField}), assuming the existence of flux surfaces $\psi$. The three constraint equations follow, and therefore it is shown that the triple product formulation of QS (\ref{QSHel}) is equivalent to (\ref{C1})-(\ref{C3}). In fact, this result does not require the assumption of magnetostatic equilibrium, as has been generally presumed, but holds under more general plasma conditions. \par

\subsection{Equivalent field line dynamics}
Given the form of $\n{u}$ in (\ref{eqn:uField}), we may rewrite (\ref{QSHel}) in the form
\begin{equation}
    \n{u}\cdot\nabla(\n{b}\cdot\nabla B)=0.
\end{equation}
Thus, $\n{u}$ may be interpreted as not only the direction of symmetry of the magnitude of $B$, but also of $\n{b}\cdot\nabla B$. The QS conditions can therefore be rewritten in the form
\begin{subequations}
\begin{gather}
     \n{u}\cdot\nabla B=0 \\
      \n{u}\cdot\nabla(\n{b}\cdot\nabla B)=0 \\
      \n{B}\times\n{u}=\nabla \psi.
\end{gather}
\end{subequations}
The physical interpretation of these equations is straightforward.\cite{Helander2014} The two scalar quantities $B$ and $\n{b}\cdot\nabla B$ govern the leading dynamics of charged particles. Thus our constraints are a statement that particles are unable to distinguish between points along $\n{u}$ that lie on different field lines but on the same flux surface. In other words, there is no way of telling field lines apart using only information about $B$ and $\n{b}\cdot\nabla B$. \par

\subsection{Comparison of constraint equations to [\onlinecite{burby2019}]}
A comparison of the three constraint equations derived in the previous section to those in [\onlinecite{burby2019}] reveals a discrepancy in the third constraint equation, which in [\onlinecite{burby2019}] is given to be
\begin{equation} 
    \n{j}\times\n{u}+\nabla(\n{u}\cdot\n{B})=0, \label{eqn:C3burby}
\end{equation}
with $\n{j} = \nabla \times \n{B}$.
In order to appreciate its origin, let us go back to Eq. (\ref{eqn:symVterm}), and rewrite it using vector identities analogous to those used in (\ref{eqn:symAterm}). As a result, 
\begin{equation}
    v_\parallel\n{\Dot{X}}\cdot\left[-\n{u}\times\n{j}+\nabla(\n{u}\cdot\n{B})\right]=0.
\end{equation}
The expression in square brackets is in fact identical to (\ref{eqn:C3burby}). However, we have shown that it is only its $\n{b}$ projection that should be made to vanish to the relevant order. Therein lies the main difference with [\onlinecite{burby2019}]: while here the approximate nature of the Lagrangian $\mathcal{L}^{(1)}$ has been taken into consideration to look for an approximate symmetry, Burby et al. sought for an exact symmetry of $\mathcal{L}^{(1)}$, which however remains an approximation to the actual physical problem. As a result, the perpendicular components of (\ref{eqn:C3burby}) lead to over-constraining the problem. To the relevant order, only the condition $\nabla\cdot\n{u}=0$ may be imposed.  \par
The conditions (\ref{C1}) - (\ref{C3}) are not only simpler but have a clear physical interpretation. Indeed, $\n{u}$ has exactly the same form as the MHD fluid flow velocity that would result from an electric field perpendicular to $\n{B}$. This is also true in other mathematical models of strongly magnetized plasmas. In neoclassical/gyrokinetic transport theory, the magnetic field is QS if, and only if, it can support such a flow of any amplitude much smaller than the sound speed \cite{Helander2014,Helander2008}.

\section{Conclusion}
In this paper, the necessary and sufficient conditions for the realisation of quasisymmetry are derived,
\begin{gather*}
    \n{u}\cdot\nabla B=0 \\
    \n{B}\times\n{u}=\nabla \psi  \\
    \nabla\cdot\n{u}=0.
\end{gather*}
where $\psi$ is a single-valued function. These conditions are obtained from the single-particle Lagrangian to the correct order. A field is to be called quasisymmetric if there exists a field $\mathbf{u}$ such that these conditions are satisfied. The Noether momentum associated with this quasisymmetry prevents particles from departing significantly from magnetic flux surfaces. \par
This set of conditions is also shown to be equivalent to other formulations of QS, such as the vanishing triple product, with the understanding that no assumption needs to be made about the magnetic field satisfying the equations of magnetostatics. It is also observed that the necessary and sufficient set of equations obtained are simpler and less constraining than those obtained by Burby et al. in [\onlinecite{burby2019}].

As Burby et al. realized, QS can be defined independently of whether the magnetic field corresponds to a plasma equilibrium. The many difficult questions concerning the existence and regularity of 3D MHD equilibria therefore do not affect our discussion although there is a close correspondence between the symmetry vector $\bf u$ and the MHD flow velocity.

\par
\hfill

\section*{Acknowledgements}
We thank the anonymous reviewers, J. Burby, N. Kallinikos and R. MacKay for constructive comments. This research is primarily supported by a grant from the Simons Foundation/SFARI (560651, AB). 

\section*{Data availability}
Data sharing is not applicable to this article as no new data were created or analyzed in this
study.

\bibliography{condQS}% Produces the bibliography via BibTeX.

\end{document}